\documentclass[aps,preprint]{revtex4}
\usepackage{graphicx}

\begin{document}

\title{Enhanced pinning and proliferation of matching effects in a superconducting film with a Penrose array of magnetic dots}

\author{A. V. Silhanek, W. Gillijns, V. V. Moshchalkov}
\affiliation{INPAC-Institute for Nanoscale Physics and
Chemistry, Nanoscale Superconductivity and Magnetism $\&$ Pulsed
Fields Group, K. U. Leuven Celestijnenlaan 200 D, B-3001 Leuven,
Belgium.}

\author{B. Y. Zhu}
\affiliation{National Laboratory for Superconductivity, Institute of Physics, Chinese Academy of Sciences, Beijing 100080, China.}

\author{J. Moonens, L.~H.~A. Leunissen}
\affiliation{IMEC vzw, Kapeldreef 75, B-3001 Leuven, Belgium.}

\date{\today}

\begin{abstract}
The vortex dynamics in superconducting films deposited on top of a
five-fold Penrose array of magnetic dots is studied by means of
transport measurements. We show that in the low pinning regime (demagnetized dots) a few periodic
and aperiodic matching features coexist. In the strong pinning regime (magnetized dots) a richer structure of unforeseen
periodic and aperiodic vortex patterns appear giving rise to a clear enhancement of the critical current in a broader field range. Possible stable vortex configurations are determined by molecular dynamics simulations.

\end{abstract}

\pacs{PACS numbers: ..............................}

\maketitle

It is a well established fact that periodic arrays of pinning centers in superconductors can substantially enhance the critical current whenever the density of vortices matches the underlying pinning landscape\cite{moshchalkov-review}. However, at high temperatures these commensurability effects occur only in a very narrow window of fields\cite{zhukov}. Moreover, for particular current orientations the periodic pinning might also promote an undesired channeling of vortices\cite{silhanek,pannetier}. In principle, these limitations could be prevented by introducing a {\it random} distribution of pinning centers at expenses of loosing the majority of the matching features\cite{silevitch,niebieskikwiat}. An alternative way to circumvent this difficulty can be achieved by using a {\it quasi-periodic} array of pinning sites as recently proposed by Misko et al.\cite{misko} On one hand, the lack of translational symmetry of these structures ensures the absolute suppression of channeling effects. On the other hand the high degree of order guarantees the persistence and even proliferation of matching features.

There is also a legitimate academic interest in the pinning properties produced by this kind of structures as the vortex pattern is forced to adopt a very peculiar configuration somehow similar to the hexatic phase in liquid crystal\cite{nelson-halperin}, with long range orientational symmetry and positional order. Besides, it is worth noticing that unlike the regular periodic lattices, in the quasi-crystal phase the vortex-vortex interaction is not perfectly compensated at every matching field thus increasing the minimum energy of the system.

In this work we determine the critical temperature $T_c$ and the critical current $I_c$ as a function of field $H$ of conventional superconductors Al and Pb with an underlying five-fold Penrose array of magnetic dots. The tunable magnetization $m$ of the dots allow us to explore two distinct regimes corresponding to weak and strong pinning as the dots are demagnetized and magnetized, respectively. We show that in the weak pinning regime matching features appear at the field $H_1=0.278 $ mT where the density of flux lines levels the density of dots, at $H/H_1=\tau$ where $\tau=(1+\sqrt{5})/2$ is the golden mean, and at $H/H_1=\nu$ with $\nu \sim 0.76$ as predicted in Ref.[\onlinecite{misko}]. When the dots are magnetized the $I_c(H)$ curve exhibits a far richer structure than in the demagnetized case with several unexpected matching features. The coexistence of many aperiodic peaks in $I_c(H)$ leads to an enhancement of the critical current over a wide field range suitable for practical applications.

The measurements were performed on a 100 nm thick Pb and a 50 nm thick Al films deposited on top of a Penrose array of magnetic dots. The dots consist of [Co(0.4 nm)/Pt(1 nm)]$_{10}$-multilayers on a 2.5 nm Pt base layer deposited by molecular beam epitaxy. Fig.~\ref{fig1}(a) shows an Atomic Force Microscopy image of the dots' array. For clarity the magnetic dots have been connected by lines to emphasize the distribution of the two types of tiles (narrow and wide) characteristic of this array. The size of the square dots is $0.7 \mu$m and the length of the connecting lines is $a = 3.1 \mu$m. The array of dots is covered with a 5 nm thick Ge layer and then a superconducting Pb or Al film is deposited on top. The Pb film has a critical temperature $T_c$ = $7.240$ K and superconducting coherence length $\xi(0)= 47.5$ nm whereas for the Al film $T_c=1.360$ K and $\xi(0)= 125$ nm. The vortex pinning properties of these hybrid samples were investigated by electrical transport measurements using a pumped He4 cryostat with base temperature of $\sim 1.1$ K.

In the as-grown state the magnetic dots have a multidomain structure with no net magnetic moment\cite{martinthesis}. Once they are magnetized by an external field it is possible to recover the original virgin state by following a strict protocol of field oscillations. Due to the small stray field present in this state, the vortex pinning is expected to be weaker than when the dots are magnetized\cite{lieve}. In order to determine the pinning efficiency of the Penrose array we measure voltage-current $V-I$ characteristics for several fields and temperatures. From these data and using a voltage criterion $V_c=3.7 \mu$V we estimate $I_c(H)$. The results of these measurements at two different temperatures for the as-grown Pb and Al samples are shown in Fig.~\ref{fig2}(a) and (b), respectively.

Assuming that the first matching field occurs when there is one flux quantum $\phi_0$ per tile, and knowing that wide tiles outnumber narrow ones by a factor of $\tau$ we find $H_1 \approx \frac{2\phi_0}{(1+\tau)a^2 sin (\pi/5)} \sim 0.279$ mT. This value agrees well with that obtained from the density of dots (0.278 mT) and the 0.28 mT estimated from the sharp reduction of $I_c$ (Fig.~\ref{fig2}). The fact that this effect is more prominent in Al than in Pb indicates that the potential well for interstitial vortices is smaller in Al probably due to a weaker random pinning. For both samples additional local increments of $I_c$ are seen at $H/H_1=\nu \sim 0.76$ and $H/H_1=\tau \sim 1.62$. The former matching condition $H/H_1=\nu$ corresponds to the occupation of all pinning sites in the vertices of the wide tiles and only three out of four in the vertices of the narrow tiles\cite{misko}. The most prominent peak above $H_1$ occurs at $H/H_1=\tau$ and corresponds to the filling of all wide rhombuses with interstitial vortices. A close inspection around $H/H_1=\nu$ suggests the existence of an additional peak near $H/H_1=1/\tau$ that gives rise to a broad plateau. This feature can be ascribed to the presence of one vortex line per wide tile. Additionally, an unexpected kink at $H/H_1\approx 0.236 \pm 0.010 \sim \frac{1}{4}$ is observed. In order to elucidate the corresponding vortex distribution for $H/H_1 \sim \frac{1}{4}$ we have performed molecular dynamics simulations of a Penrose array of Gaussian sites. Fig.~\ref{fig1}(b) shows the stable vortex configuration for $H/H_1 = \frac{1}{4}$. Here every flux line is surrounded by a corral of vacant pinning sites as indicated by the black lines. This distribution establishes certain degree of positional correlation which gives rise to the observed matching feature.

Typically commensurability effects manifest themselves at the superconductor-normal phase boundary as well. However, our determination of $T_c(H)$  using a 90$\%$ of the normal state resistance criterion (see insets of Fig.~\ref{fig2}) shows just a single feature at $H/H_1=1$ for both Al and Pb samples separating an almost linear behavior at low temperatures from a cusp close to $T_c$.

As it has been discussed in Ref.[\onlinecite{misko}] a broadening or even full suppression of matching features can occur if the elastic energy of the flux line lattice overcomes the pinning energy. This prediction suggests that by magnetizing the dots in our samples thus increasing the pinning efficiency, matching features should become easier to resolve experimentally. This is indeed the case as shown in  Fig.~\ref{fig4}(a) for the Pb sample. Now $I_c(H)$ exhibits a maximum centered at $H/H_1=1$ as a consequence of the vortex-antivortex pair generated by the stray field of the dots. At $H/H_1=1$ the external field compensates the antivortices reducing the effective field in between the dots therefore maximizing the superconducting order parameter. Unlike the demagnetized state, now two clearly resolved peaks are seen at $H/H_1-1=1/\tau$ and $\nu$. In this figure only the most obvious features have been labeled although more less sharp kinks are also visible. The observed asymmetry between $H/H_1<1$ and $H/H_1>1$ is a common behavior for all samples with magnetic dots and can be associated with the different pinning potential felt by interstitial vortices ($H/H_1>1$) and antivortices ($H/H_1<1$)\cite{martin}. A direct comparison of the field dependence of the critical current for both cases, demagnetized ($m=0$) and magnetized ($m \neq 0$) dots, is shown in the inset of Fig.~\ref{fig4}. From this figure it becomes clear that for a wide field region $I_c(H)$ is larger for the magnetized dots.

An alternative method to determine the position of the matching peaks in a more sensitive way can be achieved by using lock-in differential techniques as that obtained by higher harmonic ac-measurements\cite{zhukov} or differential resistance $dV/dI$. In order to perform $dV/dI(H)$ measurements an ac sinusoidal excitation $I_{ac}$ of frequency $f_o$ is superimposed on top of a $I_{dc}$ bias current while the ac response $V_{ac}$ is recorded. The result of these measurements is shown in Fig.~\ref{fig4}(b) for several $I_{dc}$ and $f_o=27.7$ Hz. The improved experimental resolution of the peaks allows one to identify unambiguously matching features at $H/H_1-1=1/\tau, \nu, 1$ and $\tau$. Surprisingly a very pronounced dip at $H/H_1-1=1/\tau$ with no clear counterpart at $H/H_1-1=-1/\tau$ indicates that a more stable vortex configuration is obtained at positive $H-H_1$. This can be once again related with the inherent asymmetry between vortices and antivortices\cite{martin}. Indeed, one possible scenario could be trapping of multiquanta vortices by the magnetic dots. This effect only counts for $H-H_1>0$ thus giving rise to a lower differential resistance in that field range.

To summarize, we have studied the vortex pinning properties of conventional superconductors on top of a five-fold Penrose array of magnetic dots. The possibility to tune the stray field of the magnetic dots allowed us to switch from weak to strong pinning conditions. In both limits clear aperiodic commensurability effects are identified and associated with different stable vortex configurations. The proliferation of matching peaks associated with the local symmetry of the underlying pinning array gives rise to a wide field range of enhanced critical current somehow similar to that observed in composite array of microholes\cite{composite}.

This work was supported by the Fund for Scientific Research-Flanders FWO-Vlaanderen, the Belgian Inter-University Attraction Poles IUAP, the Research Fund K.U. Leuven GOA/2004/02 and by the European ESF VORTEX programs. A.V.S. is grateful for the support from the FWO-Vlaanderen.

{\it Note added.}— While preparing this manuscript, we learned that M. Kemmler et al. at the University of Tubingen, Germany have observed similar results in samples with a Penrose array of holes in Nb superconducting films.\cite{kemmler}

\newpage


\newpage

\begin{figure}[htb]
\caption{(color online) (a) Atomic Force Microscopy image of the magnetic dots arranged in a fivefold Penrose lattice. Some dots have been linked by lines to illustrate the wide and narrow tiles. (b) Possible vortex configuration at $H/H_1 = \frac{1}{4}$ as determined by molecular dynamics simulations. Red dots indicate flux lines whereas open circles correspond to pinning sites.} \label{fig1}
\end{figure}

\begin{figure}[htb]
\centering
\caption{(color online) Critical current $I_c(H)$ for the (a) Pb and (b) Al films with the dots in the as-grown demagnetized state at two different reduced temperatures $t=T/T_c$. The insets show the corresponding phase boundaries as determined by a 90$\%$ normal resistance criterion.} \label{fig2}
\end{figure}

\begin{figure}[htb]
\centering
\caption{(color online) (a) field dependence of the critical current for the Pb film on top of a fivefold Penrose lattice of  magnetized dots. (b) differential resistance $dV/dI$ for $I_{ac}=10 \mu$A, $f_o=27.7$ Hz and several $I_{dc}$. Inset: critical current as a function of applied field for the magnetized and demagnetized dots. For clarity the horizontal axis has been normalized around the field $H_{max}$ where $I_c(H)$ maximizes.} \label{fig4}
\end{figure}

\end{document}